# Wave analysis in one dimensional structures with a wavelet finite element model and precise integration method

Shuaifang Zhang, Dongdong He, Dongsheng Li, Zhifeng Zhang, Yu Liu, Wei Shen

## ABSTRACT

Numerical simulation of ultrasonic wave propagation provides an efficient tool for crack identification in structures, while it requires a high resolution and expensive time calculation cost in both time integration and spatial discretization. Wavelet finite element model provides a high-order finite element model and gives a higher accuracy on spatial discretization, B-Spline wavelet interval (BSWI) has been proved to be one of the most commonly used wavelet finite element model with the advantage of getting the same accuracy but with fewer element so that the calculation cost is much lower than traditional finite element method and other high-order element methods. Precise Integration Method provides a higher resolution in time integration and has been proved to be a stable time integration method with a much lower cut-off error for same and even smaller time step. In this paper, a wavelet finite element model combined with precise integration method is presented for the numerical simulation of ultrasonic wave propagation and crack identification in 1D structures. Firstly, the wavelet finite element based on BSWI is constructed for rod and beam structures. Then Precise Integrated Method is introduced with application for the wave propagation in 1D structures. Finally, numerical examples of ultrasonic wave propagation in rod and beam structures are conducted for verification. Moreover, crack identification in both rod and beam structures are studied based on the new model.

## Introduction

The safety, durability, and sustainability are essential requirements for civil structures in the long-termservice. The performance and the safety of civil structures would be gradually weakened due to possible damage from artificial factors or natural disasters. Thus, damage detection have attracted plenty of attention in different industry areas, i..e, aerospace, mechanical, materials science and civil engineering {Cha 2015 & 2017; Cole, 2004; Song, 2006; Song, 2010; Wandowski 2015; Zhang, 2017, Cheng}. The effective detection methods have been proved to be an effective way to prevent structure failure. Different approaches have been developed for damage detection in structures, such as vibration-based damage detection approaches{Yam, 2003;Chen, 2007 ;Adewuyi, 2011;Bagheri, 2009}, test method{An, 2015}, model-free method {Li, 2008}, wrapping method{Nicknam, 2011}, axial strain method{Blachowski, 2017}, deep response surface method{Zhou, 2013}, learning method{Cha, 2017} and so on. Nondestructive testing to detect small damages is still a challenging problem since it requires very high-frequency excitation guided wave signal. Numerical simulation of guided wave propagation in structures has been studied for a long time and numerous models has been applied for the simulation of guided wave propagation. A lot of numerical models have been developed for simulation of wave propagation in 1D structures{Kudela, 2007; Harari, 1995, Anderson, 2006; Seemann, 1996; Xiang, 2007}. Finite Element Method{Marfurt, 1984; Moser, 1999} is most widely used method because of its adaptivity for different shapes and dimensions in structure. However, it requires very fine mesh for structure, i.e., at least ten nodes per wavelength are required to obtain an accurate simulation result{Chen, 2012}. Hence the element size should be small enough for ultrasonic waves with high frequencies in FEM-based simulation. To improve the computation efficiency, wave propagation problems are usually solved in frequency domain and then convert the solution to time domain. FFT based spectral element method{Doyle, 1989} is one solution because it could transform the governing PDEs to a set of ODEs, which are much easier to solve, with constant coefficients. Then the solution of ODEs in frequency domain is transformed into time domain with inverse transform (IFFT). However, FFT based SEM is only applicable to solve problems of infinite rods, semi-finite rods, and beams due to the assumption of the periodic FFT. Laplace transform is subsequently applied by different researchers to replace FFT to overcome such problems{Igawa, 1999}. Thus, the time domain SEM{Kudela, 2007} is much more effective than FFT based spectral element method, especially in guided wave nondestructive testing with small damages, because each element in SEM has more than one node and could get the same accuracy with fewer elements.

Wavelet transform analysis has been widely used in engineering applications due to its time-frequency analysis features, wavelet-based finite element method has also got its great success in modeling guided wave propagation. There are also two different kinds of wavelet finite element methods, one of which is time domain wavelet finite element method { Ma, 2003; Xiang, 2007; Amiri, 2015; Chen, 2012}, which is very similar with the traditional finite element method in time domain but with high spatial resolution, while the other one is frequency domain finite element method{Mitra, 2005}. Both methods have been applied in numerical simulation of guided wave propagation and crack identification in structures, time domain wavelet finite element method has been widely used since its feature of wide adaptivity in space, another advantage is that it requires much less DOFs (almost save 75% of DOFs space). However, the small time interval is still required, so the computational cost for this method is still expensive.

As we know that the element type plays a very important role in getting accurate results and improving the efficiency, and almost all these research papers focus on the element type instead of time integration method. In a consequence, the high-order elements such as Spectral element and Wavelet finite element have been widely used in numerical simulation of ultrasonic guided wave propagation in structures. It's also very important for us to study the time integration method and find an efficient and stable integration method. Newmark-beta method has been widely used as time integration method due to its stable property. Recently, a Fourier transform based wavelet finite element model{Shen, 2017} has got great success due to its low requirement on the time integration step, but it still has the limitation that it requires the time samples are large enough to find the accurate solution. Precise integration method, which is proposed by Zhong, has been widely used in highly nonlinear heat transfer problems and vibrations due to its high stability and high accuracy. More importantly, it could always provide the precise numerical solution for certain cases, especially for ultrasonic guided waves with known excitation forces. So it's also very important for guided wave propagation researchers to know this important work.

This manuscript provides the wavelet finite element model combined with precise integration method, as wavelet finite element model provides high resolution in space coordinate while precise integration method will need very small time integration step to get the accurate results. Chapter two will introduce how to construct the wavelet finite element model for 1D structures; then chapter three will give a brief introduction on the precise integration method. Chapter four will provide numerical examples of ultrasonic guided wave propagation in 1D structures. Also the time integration step will be discussed for verification of the advantages of our model and method; Finally, crack identification in rod and beam structures will be presented for validation.

# 2. Numerical models

## 2.1 B-spline wavelet on interval finite element

Unlike Fourier transform, wavelet transform has two different variables, scale $a$ and translation $\tau$, and the wavelet transform is defined as:

$$W(a,\tau) = \frac{1}{\sqrt{a}} \int f(t) * \psi(\frac{t-\tau}{a}) dt$$

Where $\psi(t)$ is a wavelet function. There is another important function called scaling function $\phi$ in wavelet transform. The scaling function is used to build the spaces in multi-resolution spaces:

$$V_j = Span\{\phi_{j,k}, j, k \in Z\}$$

There are many wavelet and scaling functions developed by researchers and engineers, in this manuscript, B-Spline scaling functions will be used to construct the wavelet finite element model. 0 scale $m$th order B-spline scaling function and wavelet function are developed by Goswami [20]. The 0 scale 2$^{nd}$ order scaling functions ($j=0$, $m=2$) is expressed as:

$$\phi^0_{2,-1}(\xi) = \begin{cases} 1-\xi, \xi \in [0,1] \\ 0, \; else \end{cases} \tag{3}$$

$$\phi^0_{2,0}(\xi) = \begin{cases} 1-\xi, \; \xi \in [0,1] \\ 2-\xi, \; \xi \in [1,2] \\ 0, \; else \end{cases} \tag{4}$$

All other scaling functions could be built based on the definition, here we take $\phi^3_{4,k}(\xi), (k = -3,-2,\ldots,7)$ to build destinated B-Spline wavelet on interval finite element model, specifically, the plot for the scaling functions for 4$^{th}$ order and 3$^{rd}$ scale are shown in Fig. 1.

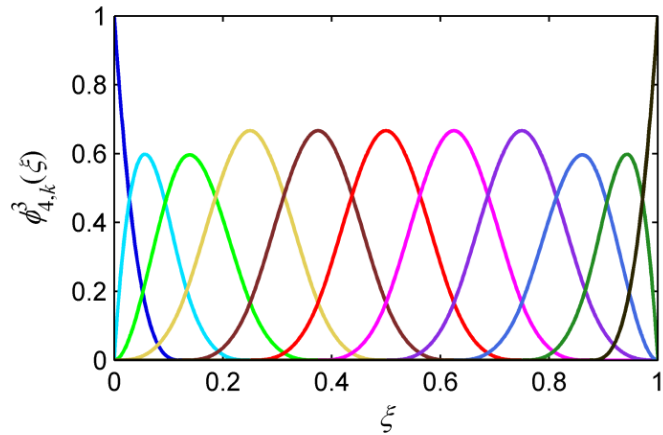

Fig.1. Plot of B-Spline scaling functions

The Wavelet finite element model for ultrasonic wave propagation is the same as the general finite element model:

$$M\ddot{\boldsymbol{u}}(t) + K\boldsymbol{u}(t) = f(t)$$

Where $M$ is the mass matrix, $K$ is the stiffness matrix, while $\boldsymbol{u}(t)$ is displacement as a function of time, $f(t)$ is the load. Here we will show a brief introduction of the derivation of the mass matrix and stiffness matrix.

Firstly, the nodal displacement could be expressed as a sum of the scaling functions and its coefficients.

$$u(\xi) = \sum_{k=-m+1}^{2^j-1} a_{m,k}^j \phi_{m,k}^j(\xi) = \sum_{k=-3}^{7} a_{4,k}^3 \phi_{4,k}^3(\xi) = \boldsymbol{\Phi} \boldsymbol{a}^e$$

Where $\boldsymbol{\Phi}$ is a row vector with 11 scaling functions, and $\boldsymbol{a}^e$ is a column vector of 11 wavelet coefficients. For one element with 11 nodes, the nodal displacement matrix $\boldsymbol{u}^e$ could be expressed as:

$$\boldsymbol{u}^e = \boldsymbol{\Phi} \boldsymbol{a}^e = [\boldsymbol{R}^e]^{-1} \boldsymbol{a}^e$$

Where $\boldsymbol{u}^e = [u(\xi_1)\, u(\xi_2)\, \cdots\, u(\xi_n)]^{\mathrm{T}}$ is the nodal displacement vector for the nodes in one element, $[\boldsymbol{R}^e]^{-1} = \boldsymbol{\Phi} = [\boldsymbol{\Phi}^T(\xi_1)\, \boldsymbol{\Phi}^T(\xi_2)\, \cdots\, \boldsymbol{\Phi}^T(\xi_{n+1})]$, by substituting the solution of $\boldsymbol{a}^e$, we could get the displacement as a function of nodal DOFs,

$$u(\xi) = \boldsymbol{\Phi} \boldsymbol{a}^e = \boldsymbol{\Phi} \boldsymbol{R}^e \boldsymbol{u}^e = \boldsymbol{N}^e \boldsymbol{u}^e$$

In which $\boldsymbol{N}^e = \boldsymbol{\Phi} \boldsymbol{R}^e$ is the shape function. There are 11 shape functions in total, for convenience, here we would like to show the plots of the shape functions:

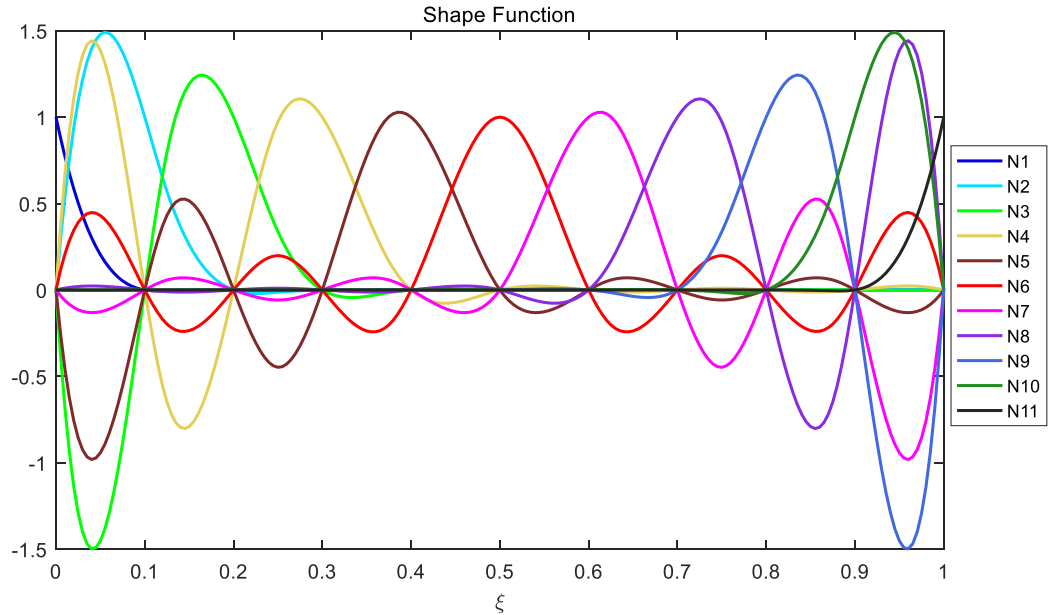

Fig.2 Shape functions for BSWI FEM

From the plots of different shape functions for BSWI FEM, there is much difference for BSWI FEM than traditional FEM that shape functions of BSWI FEM may be larger than 1. But what is the same as traditional FEM is that both satisfies a basic property:

$$N_i(\xi_j) = \delta_{ij} \tag{9}$$

## 2.2 Numerical model in 1D structure

The most common 1D structures are rod structure and beam structure, for rod structure, there is only 1 DOF, and the displacement is expressed in the way shown previously, based on the displacement we get, we could find the potential energy and kinetic energy first,

$$U^e = \int \frac{EA}{2}\left(\frac{\partial u}{\partial x}\right)^2 d\xi \tag{10}$$

$$T^e = \int \frac{\rho A}{2}\left(\frac{\partial u}{\partial t}\right)^2 d\xi \tag{11}$$

Where E is the Young's modulus, A is the cross section area, $\rho$ is the density. Then the mass matrix and stiffness matrix of BSWI rod element could be obtained based on Hamilton's variation principle,

$$M = \rho A l_e \int [N^e]^T N^e d\xi = \rho A l_e \int [R^e]^T [\Phi]^T \Phi R^e d\xi$$

$$K^e = \frac{EA}{l_e} \int [\frac{dN^e}{d\xi}]^T \left[\frac{dN^e}{d\xi}\right] d\xi = \frac{EA}{l_e} \int [R^e]^T [\frac{d\Phi}{d\xi}]^T [\frac{d\Phi}{d\xi}][R^e] d\xi$$

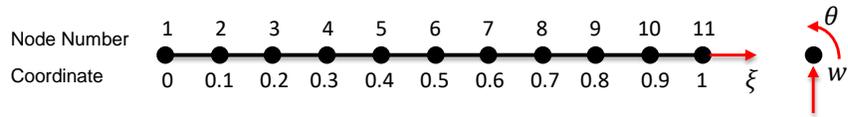

Fig. 1 The basic BSWI43 beam element

Each node on beam structures has 2 DOFs based on the assumption of Timoshenko beam, the deflection $w$ and the angle of rotation $\theta$. Both DOFs could be expressed as a function of nodal displacement/rotation and shape functions:

$$w = N^e w^e = \Phi R^e w^e \tag{9}$$

$$\theta = N^e \theta^e = \Phi R^e \theta^e \tag{10}$$

Again, based on Hamilton's Principle, we could find the mass and stiffness matrix for one BSWI element:

$$K^e = \begin{bmatrix} K_1^e & K_2^e \\ K_3^e & K_4^e \end{bmatrix} \tag{12}$$

$$M^e = \begin{bmatrix} M_1^e & 0 \\ 0 & M_2^e \end{bmatrix} \tag{13}$$

Where, $K_1^e = \frac{GA}{kl_e}\int[R^e]^T[\frac{d\Phi}{d\xi}]^T[\frac{d\Phi}{d\xi}][R^e]d\xi$, $K_1^e = (K_2^e)^T = -\frac{GA}{k}\int[R^e]^T[\frac{d\Phi}{d\xi}]^T[\Phi][R^e]d\xi$

$$K_4^e = \frac{EI}{kl_e}\int[R^e]^T[\frac{d\Phi}{d\xi}]^T\left[\frac{d\Phi}{d\xi}\right][R^e]d\xi + \frac{GAl_e}{k}\int[R^e]^T[\Phi]^T[\Phi][R^e]d\xi$$

$$M_1^e = \rho A l_e \int [R^e]^T [\Phi]^T \Phi R^e d\xi$$

$$M_2^e = \rho I_y l_e \int [R^e]^T [\Phi]^T \Phi R^e d\xi$$

## 2.3 Crack model in 1D structures

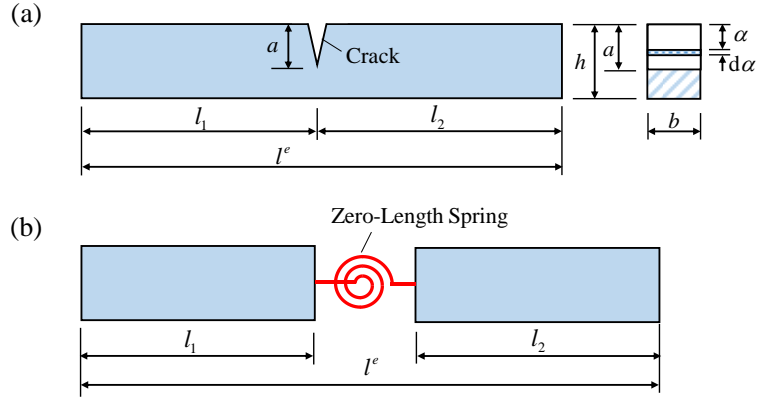

Fig. 4 The crack model in rod and beam element: (a) a cracked element, (b) spring model[Shen]

There are many crack models, Fig. 5 shows the most popular crack models in rod and beam structures used by most researchers. The axial flexibility of spring could be calculated as:

$$c_a = \frac{2\pi}{Eb} \int_0^a \frac{\alpha}{h^2} f_I^2(\alpha) \, d\alpha \tag{21}$$

Where $\alpha$ is the current location coordinate, $E$ is the Young's modulus; $b$ and $h$ are the width and height of cross-section, respectively

$$f_I(\alpha) = \sqrt{\frac{\tan(\pi\alpha/2h)}{\pi\alpha/2h}} \times \frac{0.752 + 2.02\alpha/h + 0.37[1-\sin(\pi\alpha/2h)]^3}{\cos(\pi\alpha/2h)} \tag{22}$$

For beam structures, rotational stiffness and shearing stiffness are used to stimulate the crack in Timoshenko beam, the rotational flexibility $c_b$ and the shearing flexibility $c_s$ are expressed as:

$$c_b = \frac{72\pi}{Ebh^2} \int_0^a \frac{\alpha}{h^2} f_I^2(\alpha) \, d\alpha \tag{21}$$

$$c_s = \frac{2k^2\pi}{Eb} \int_0^a \frac{\alpha}{h^2} f_{II}^2(\alpha) \, d\alpha \tag{22}$$

Where $\alpha$ is the current location coordinate, $E$ is the Young's modulus; $b$ and $h$ are the width and height of cross-section, respectively

$$f_I(\alpha) = \sqrt{\frac{\tan(\pi\alpha/2h)}{\pi\alpha/2h}} \times \frac{0.752 + 2.02\alpha/h + 0.37[1-\sin(\pi\alpha/2h)]^3}{\cos(\pi\alpha/2h)} \tag{23}$$

$$f_I(\alpha) = \sqrt{\frac{\tan(\pi\alpha/2h)}{\pi\alpha/2h}} \times \frac{0.752 + 2.02\alpha/h + 0.37[1-\sin(\pi\alpha/2h)]^3}{\cos(\pi\alpha/2h)} \tag{24}$$

# 3. Precise Integration Method

To solve the ultrasonic wave propagation problem, Newmark, Wilson-θ and Runge-Kutta are the most common time-domain integral methods. However, the integral step size of these methods has to be small enough for accurate results due to the requirement of stability and accuracy, thus the calculation cost is very expensive. In this paper, the PIM presented by Zhong and Williams [1], which allows for a longer time step for integration and offers the advantages of high accuracy and good stability [2], is used to solve the linear system accurately and efficiently.

In state space, Eq. (5) can be written as a set of ODE functions in a matrix form:

$$\dot{z} = Hz + F(t)$$

Where

$$z = \{u \quad \dot{u}\}^T$$

$$H = \begin{bmatrix} 0 & I \\ -M^{-1}K & 0 \end{bmatrix}, F(t) = \begin{bmatrix} 0 \\ -M^{-1}f(t) \end{bmatrix}$$

To solve the ODE functions in state space, we could solve this step by step for Homogeneous and nonhomogeneous differential equations according to the similar steps in ODE.

As we know that the solution of ODE function $\frac{dz}{dt} = \lambda z$, with initial condition $z(0) = z_0$ is:

$$z(t) = z_0 e^{\lambda t}$$

Similarly, for a set of ODE functions in matrix form $\frac{dz}{dt} = Hz$, with initial condition $z(0) = z_0$, where $z$ is a vector with n elements $z = \{z_1, z_2, \ldots, z_n\}^T$, and $H$ is a n dimensional matrix. We could find the solution

$$z = e^{H\tau} z_0$$

Suppose the time integration step is $\tau$ and a very big number $M = 2^N$, initial time step $t_0 = 0, t_1 = \tau$, and the kth time step $t_k = k\tau$, also suppose $\tau = M\Delta t = 2^N \Delta t$, thus we could find the solution at different time steps:

$$z_1 = e^{H\tau} z_0, z_2 = e^{H\tau} z_1, \ldots, z_k = e^{H\tau} z_{k-1}$$

Suppose $T = e^{H\tau} = \left[e^{H\tau/M}\right]^M = \left[e^{H\Delta t}\right]^M$, as M is a very large number, we could get the following equation holds:

$$e^{H\Delta t} = I + A\tau + \frac{1}{2!}(A\tau)^2 + \frac{1}{3!}(A\tau)^3 + \cdots = I + T_1$$

And $T = e^{H\tau} = \left[e^{H\Delta t}\right]^M = [I + T_1]^M = [I + T_1]^{2^N} = I + I_{N+1}$

After obtaining the matrix **T,** for the (k+1)th time step $t \in [t_k, t_k + \tau]$, the solution of the state function could be found by Duhamel integration as

$$z_{k+1} = Tz_k + \int_0^\tau e^{H(\eta - \xi)} F(t_k + \xi) d\xi$$

For common loads, the integral in Eq. 错误!未找到引用源。 can usually be solved analytically. If the excitation is harmonic within a time step, i.e.,

$$\mathbf{F}(t) = \mathbf{r}_1 \sin(\omega t) + \mathbf{r}_2 \cos(\omega t) \qquad (1)$$

Eq. 错误!未找到引用源。 can be written as

$$\mathbf{z}_{k+1} = \mathbf{T}\left[\mathbf{z}_k + \mathbf{C}_1 \sin(\omega t_k) + \mathbf{C}_2 \cos(\omega t_k)\right] - \mathbf{C}_1 \sin(\omega t_{k+1}) - \mathbf{C}_2 \cos(\omega t_{k+1}) \qquad (2)$$

where

$$\begin{aligned}\mathbf{C}_1 &= \left(\omega^2 \mathbf{I} + \mathbf{H}^2\right)^{-1}\left(\mathbf{H}\mathbf{r}_1 - \omega \mathbf{r}_2\right)\\ \mathbf{C}_2 &= \left(\omega^2 \mathbf{I} + \mathbf{H}^2\right)^{-1}\left(\mathbf{H}\mathbf{r}_2 + \omega \mathbf{r}_1\right)\end{aligned} \qquad (3)$$

In Eq. (2), the matrix exponential $\mathbf{T}$ can be computed accurately by using the PIM [1], and $\mathbf{C}_1$ and $\mathbf{C}_1$ can be computed accurately easily by solving Eq. (3).

# 4. Numerical examples of ultrasonic wave propagation in 1D structure

## 4.1 Time interval influence study of ultrasonic wave propagation in rod structure

An excitation signal with 5-cycle sinusoidal tone burst is picked for wave propagation simulation in rod structure, the single central frequency of which is 100kHz. The excitation signal in time domain is listed in Eq. 23.

$$f(t) = \begin{cases} \frac{1}{2}[1 - \cos(40\pi t)] \sin(200\pi t) & 0 \leq t \leq 0.05ms \\ 0 & others \end{cases} \qquad (23)$$

Also, the plots of the excitation signal in time domain and wavelet time-frequency spectrum are shown in Fig. 5.

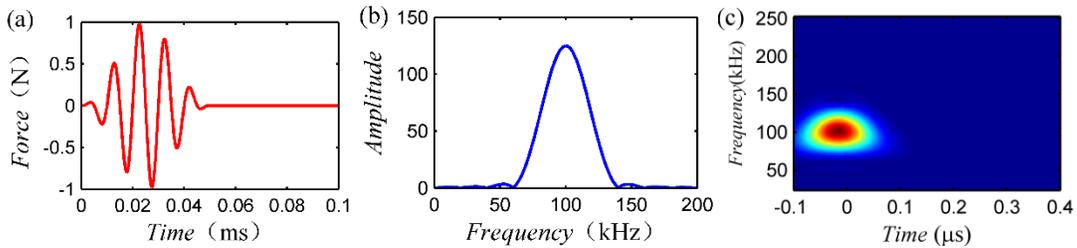

Fig. 5 Excitation signal : (a) Time-domain diagram; (b) Frequency spectrum; (c) Wavelet time-frequency spectrum

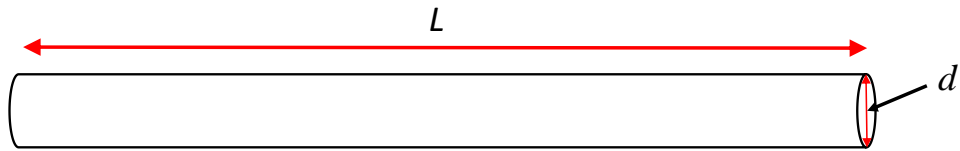

Fig. 6. Schematic plot for rod structure

Here, we take an Aluminum rod structure with length L=1500mm and the diameter d=12mm, and the excitation signal load is applied on the left end.

| Length (mm) | Young's Modulus (GPa) | Poisson's ratio | Density (kg/m3) | Diameter(mm) |
|---|---|---|---|---|
| 1500 | 70 | 0.3 | 2730 | 12 |

Tab. 2 Material properties and geometry parameters of Al rod

Firstly, in order to study the influence of time interval influence on the simulation results, different time interval and same element size are applied for study, the minimum time interval is set as $\Delta t = T_{min}/20 = 0.33\mu s$ to ensure 20 integration time steps per period, and we compared with several other cases with $\Delta t = 5\mu s, 10\mu s, 50\mu s$, the reason why we pick these three time interval is because that the excitation signal period is $50\mu s$. And the signal received by the right end is shown below:

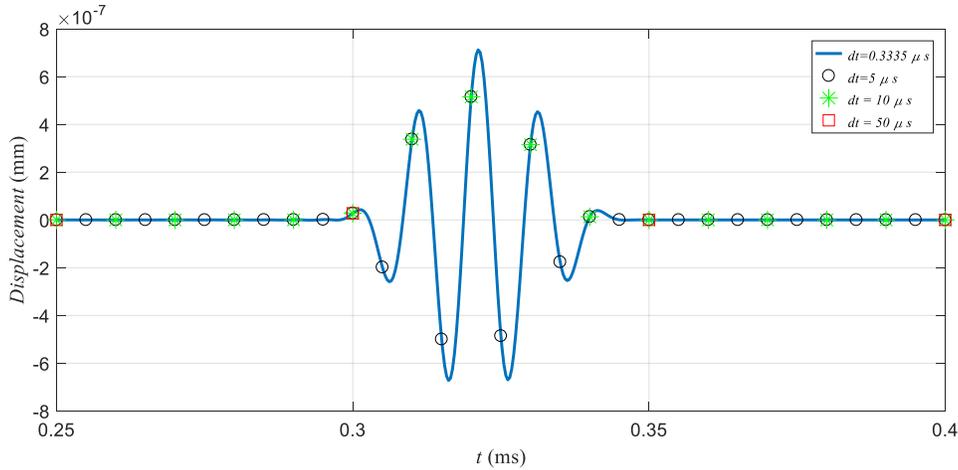

Fig.7 Displacement at the right end of the rod for different time intervals

As we could see from the figure, with the increasing the of time interval, the sample points we get for the displacement signal are less and less. However, all the points are still on the displacement curve, which means the results show the correct solution for even $\Delta t = 50\mu s$. For $\Delta t > 50\mu s$, the solution will not be correct any more since the displacement is always zero and the period of the excitation signal is only $50\mu s$. Thus, for a time integration method, the maximum time interval for this problem is $50\mu s$, and we have proved that for the maximum time interval we could always get the accurate answer, and this time interval is much larger than the time interval got in Fourier transform based wavelet finite element method in our previous paper published in JSV. For a better proof that the results we got are correct, we present the nodal displacement along the rod at different time steps:

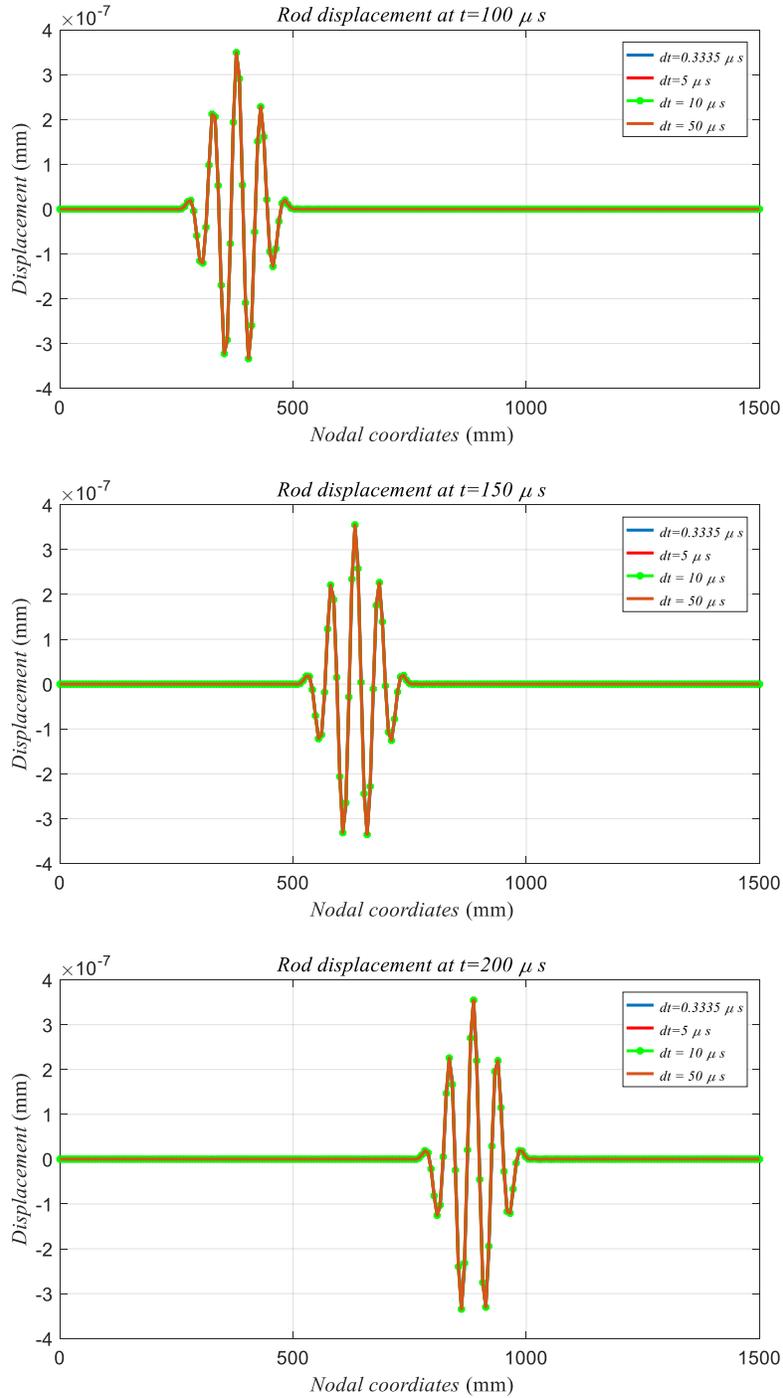

Fig.8 Nodal displacement along the rod at different time steps

As shown in the figure, the nodal displacement for different time intervals are exactly the same, which indicates even the largest time interval shows the correct solution of the displacement along the rod structure.

As from the reference, we could see that almost the most efficient model right now in published papers is FFT based wavelet finite element model, and minimum time interval in that paper shown is 3.3 $\mu s$, which has improved 10 times compared with BSWI FEM. However, with Precise Integration method, the minimum time interval could be as small as 50 $\mu s$, which is almost 150 times of the time interval required for BSWI FEM, which is much faster than conventional FEM. Detailed information is shown in Tab. 3.

| Different methods | Number of elements | Number of DOFs | Time interval |
|---|---|---|---|
| **Conventional FEM** | 712 | 713 | 0.33$\mu s$ |
| **BSWI FEM** | 16 | 161 | 0.33$\mu s$ |
| **FFT-based BSWI** | 16 | 161 | 3.3$\mu s$ |
| **PIM-based BSWI** | 16 | 161 | 50$\mu s$ |

Tab. 3 Spatial discretization and time interval requirement comparison for different methods with same length of rod in reference[Shen]

### 4.2 Ultrasonic wave propagation in rod structures

Here we will take the maximum time interval for calculation, and the size of the rod is the same as the previous chapter. For comparison with the accurate result, we would like to find the velocity of the ultrasonic wave propagation in rod structure. As the theoretical velocity could be calculated as $c_0 = \sqrt{E/\rho} = 5063 m/s$. As the time interval we picked is $\Delta t = 50\mu s$, so we compared the results at different time steps for the first 9 time steps for comparison, which is shown in the following figure.

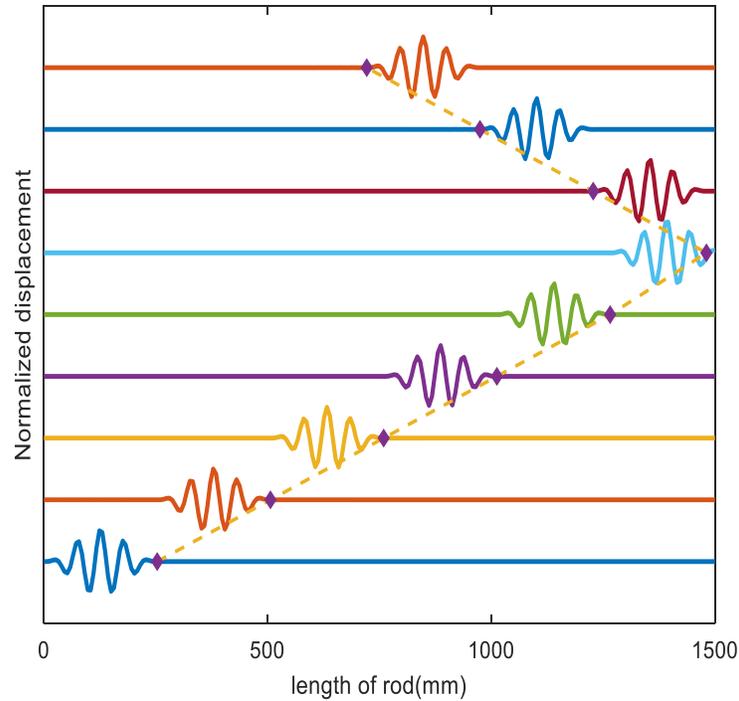

Fig. 9 Normalized displacement along rod structure at different time steps with $\Delta t = 50\mu s$

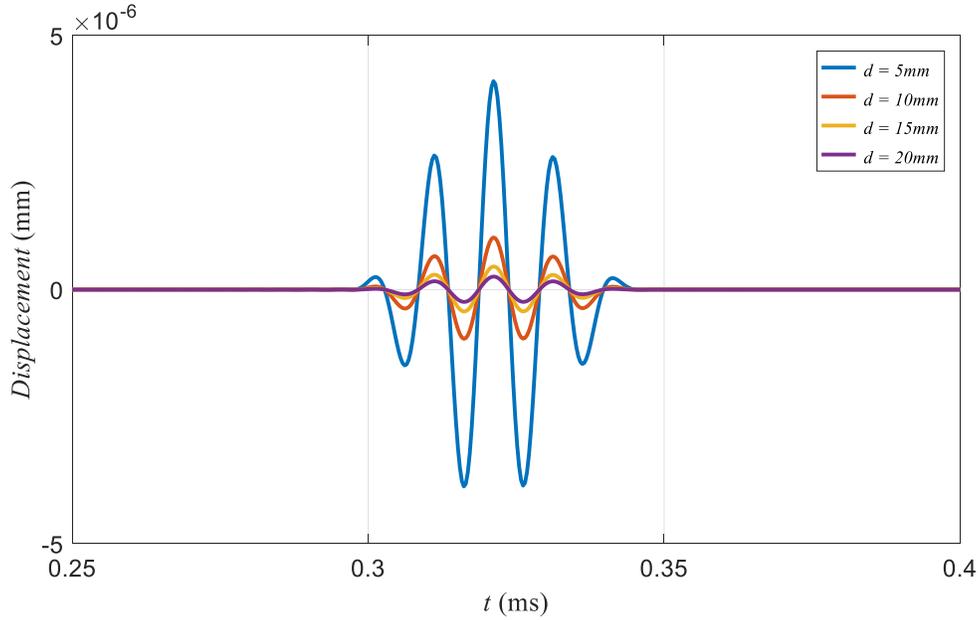

Fig. 10 Displacement curve in rod with different diameters

Then we studied the influence of diameters on the wave propagation in rod structures. Fig.10 shows the displacement signal received at the right end of the rod with different diameters. Firstly, we could see that the arrival time for different rod diameters are almost the same, which means the group velocity for the ultrasonic guided wave with the same frequency is the same, apparently this proves that the group velocity is not influenced by the diameter of the rod structures. Secondly, the figure also shows that the amplitudes of the wave signals are different, and the amplitude would decrease if the diameter is increasing.

## 4.3 Ultrasonic wave propagation in beam structures

In this part, ultrasonic wave propagation in beam structures will be studied, steel is chosen as the material for the beam structure. The specific properties of steels are different because of processing and compositions {Cheng, 2016& 2017}. The detailed material property and geometry parameters used in this article are shown in Table 3.

| Length (mm) | Young's Modulus (GPa) | Poisson's ratio | Density (kg/m3) | Width(mm) | Height(mm) |
|---|---|---|---|---|---|
| 1800 | 200 | 0.3 | 7800 | 12 | 12 |

Table 3. Material property and geometry parameter of steel beam

Since each node has 2DOFs in beam structures, here we will study the displacement behavior on both DOFs at the right end of the beam. The excitation signal is loaded on the left end, and the response signal for both DOFs at the right end of the beam are shown in Fig. 11.

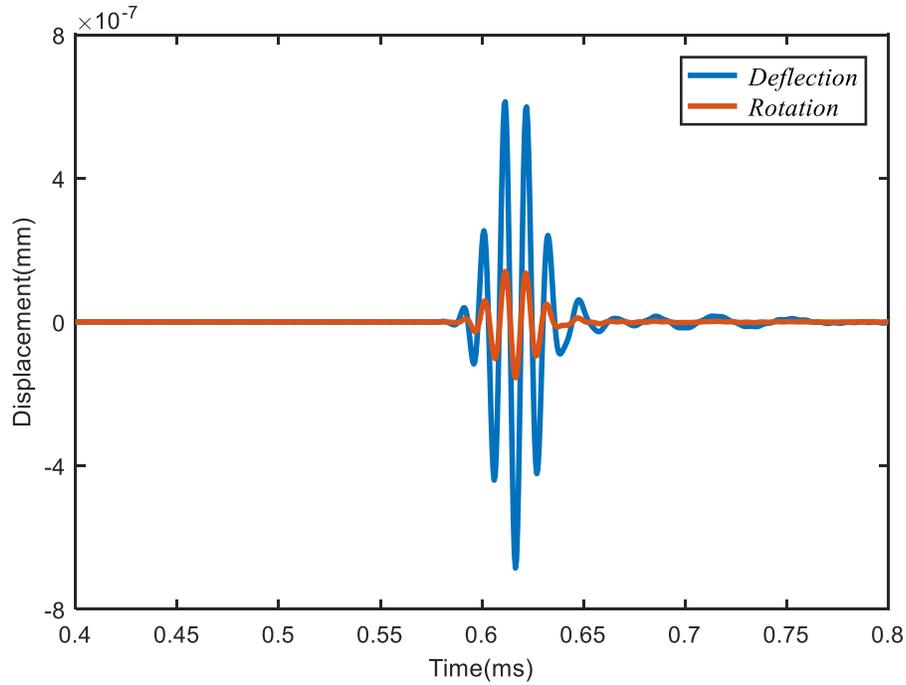

Fig.11. Displacement response at right end of beam

As shown in Fig.11, the deflection and rotation in beam structure show a bit different. Firstly, the amplitude of deflection is much larger than the amplitude of rotation; secondly, though the waveforms are almost the same, there is still slight difference after the signal arrived. Besides, from the figure it is observed that the arrival time for both DOFs is almost the same, this is because both DOFs correspond to one same node.

Figure 12 shows the deflection and rotation response along beam at different time steps with a time difference of 100ms for adjacent steps, also the time frequency plots for the deflection response is shown in the figure. As shown in Fig. 12, the normalized deflection and rotation response travels along the beam with almost the same constant velocity.

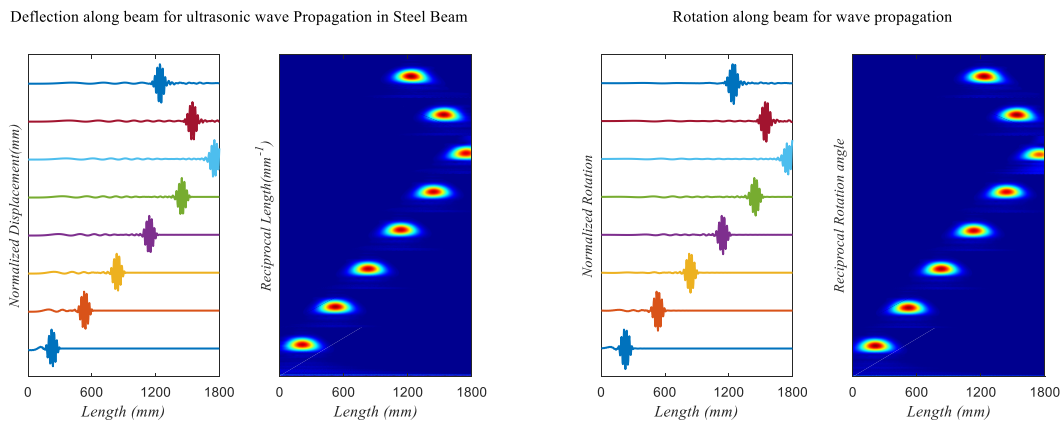

Fig. 12. Normalized and Time frequency analysis of deflection and rotation response along beam at different time

## 4.4. Crack detection with BSWI finite element and precise integration method in 1D structures

Firstly, a numerical example for ultrasonic wave propagation in rod is studied, the material is still Aluminum rod, the length of the rod is set as 1500mm. Instead of analyzing a round rod, the section is assumed as a 12mm square. Suppose there is a crack in the middle of rod, where the crack depth is set as 20% of the height, to determine the location of crack, one sensor is enough for crack detection, which is set on the right end of the rod. And the received signal is shown below:

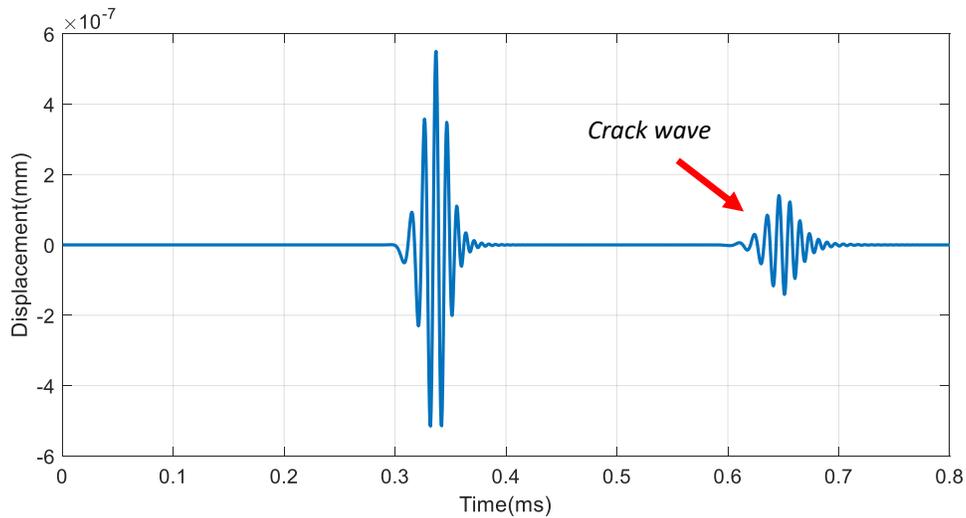

Fig. 13 Signal received by the right end of rod with a crack in the middle

Shen has studied the influence of crack location on the wave signal received by the right end, which shows that when the crack is in the middle of the rod, the amplitude of the crack will be larger than the other crack locations [reference]. To investigate and prove this phenomenon, different locations of crack in the rod are studied, due to the symmetry of the rod, 5 different crack locations are studied and all of them are located at the left half of the rod. Specifically, the cracks are set at the locations: $0.1l = 150mm, 0.2l = 300mm, 0.3l = 450mm, 0.4l = 600mm, 0.5l = 750mm$, respectively. And the signals received by the right end are shown in Fig. 14.

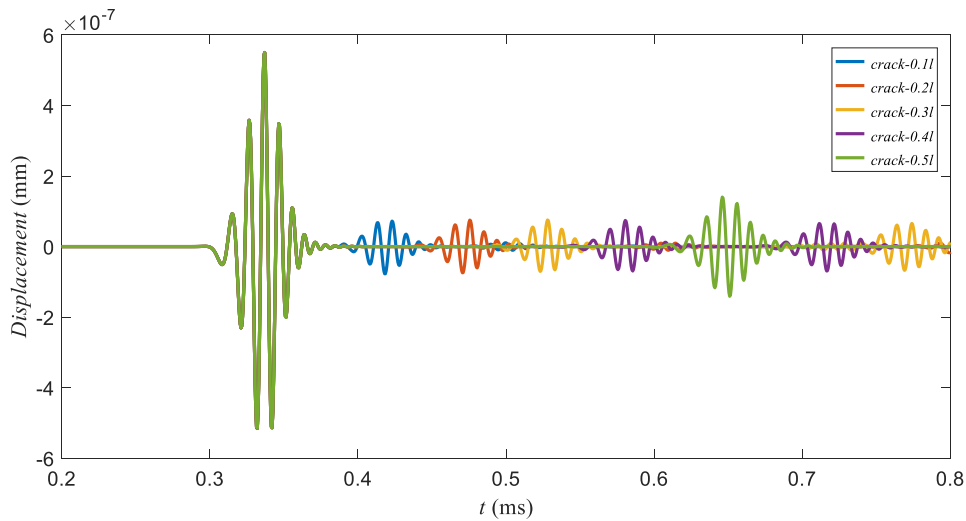

Fig. 14. Different crack locations influence on the received signal at the right end of rod

As shown in Fig. 14, the crack wave received by the right end with the middle crack has the largest crack wave amplitude. Besides, the direct waves received for different crack locations are the same, apparently this also proves that precise integration method is a reliable method.

*Wave Propagation in rod with different crack locatioins*

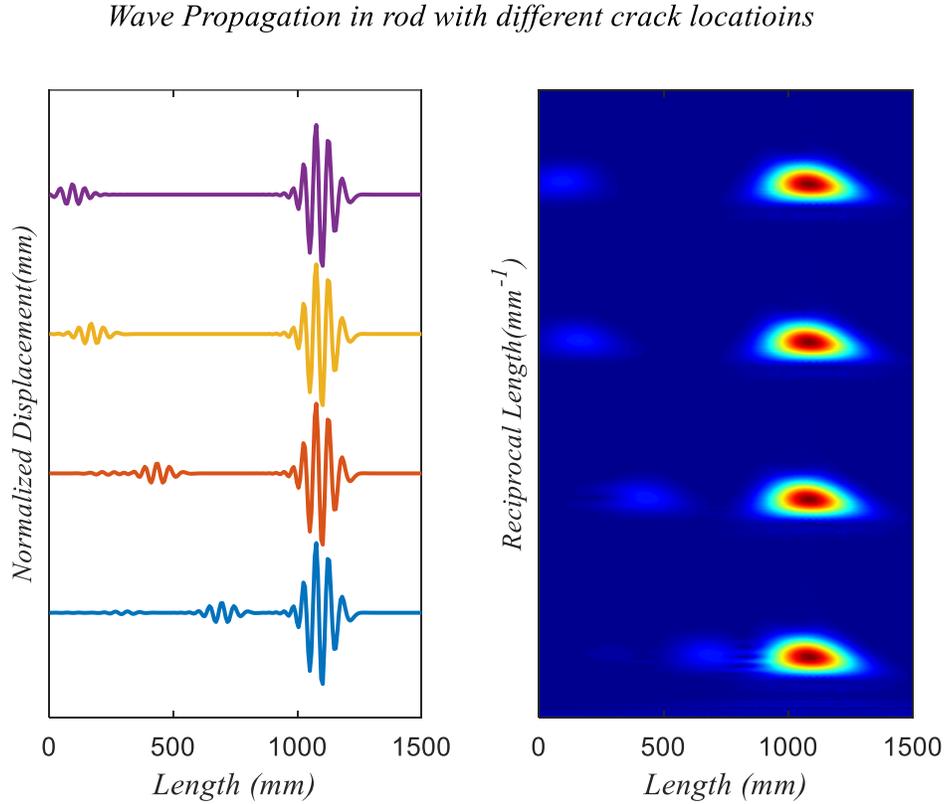

Fig. 15 Time frequency analysis of wave propagation in rod with different crack locations

Fig.15 shows the wave signal received at $t = 250 \mu s$ with different crack locations, where the cracks are located at $0.1l, 0.2l, 0.3l, 0.4l$ from bottom to top, respectively. From the time frequency plot, we could see that the direct wave arrives at the same location while the crack waves arrive at different locations along the rod. It's obvious that the crack wave arrives earlier when the crack location is closer to the left end of the rod.

Secondly, a beam with length of $1000\ mm$ is presented, the crack is located in the middle of the beam, the material is made of Aluminum, where the detailed material properties are shown in Tab. 2 in the previous chapter. The signal received by the right end of the beam is shown in Fig. 16. Apparently, we could see that PIM based BSWI FEM could accurately solve ultrasonic wave propagation in beam structures. Fig. 17 shows the displacement signal received at time $t = 250\ \mu s$, where deflection is picked for representation and nodal deflection along the beam DOFs. Also, the crack wave propagation is behind the direct wave propagation, which is quite reasonable.

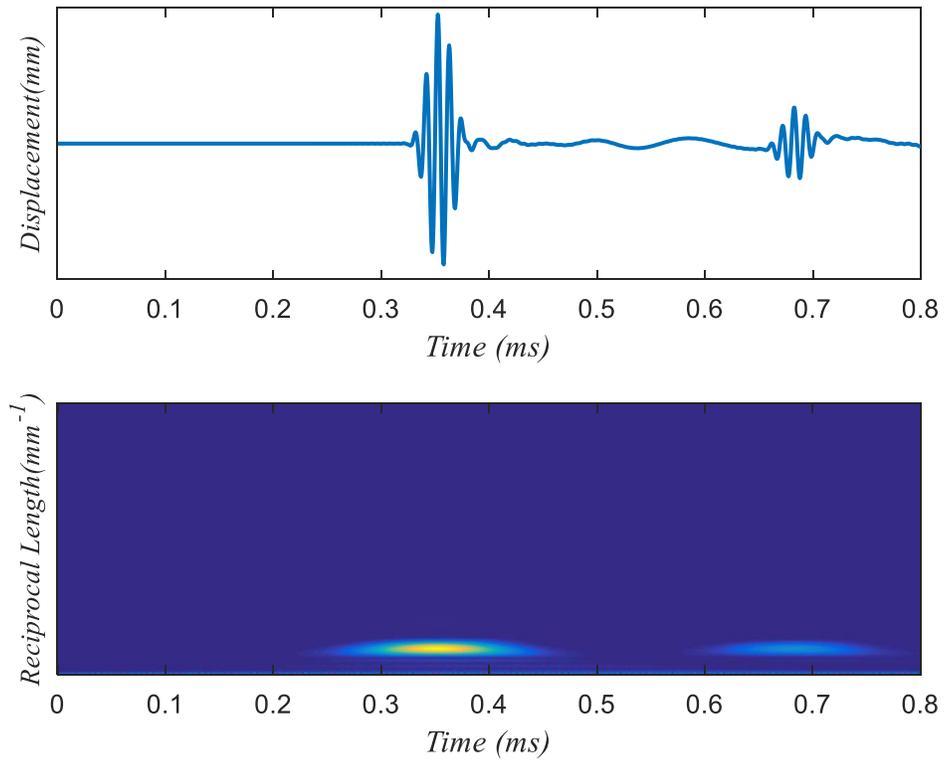

Fig. 16. Signal received by the right end of the beam with crack in the middle

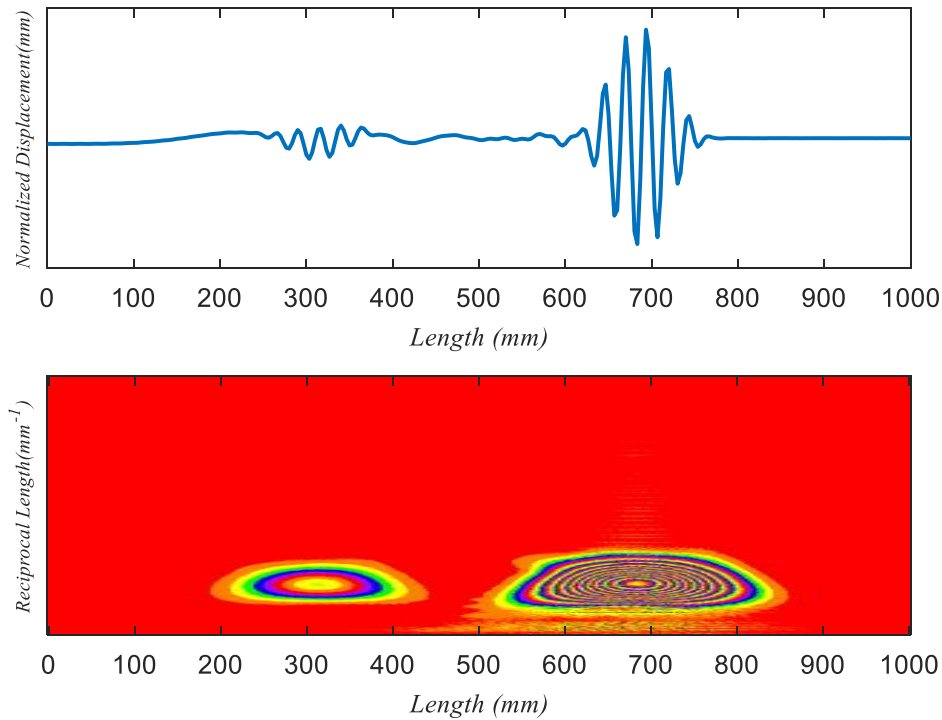

Fig.17. Deflection displacement along beam under time $t = 250\mu s$

## 5. Conclusion

This paper presents a BSWI finite element method to solve ultrasonic wave propagation problems in 1D structures, as ultrasonic wave has the feature of high frequency and it requires a very small time integration step, so this paper combines a new and precise integration method to solve this problem. Following conclusions could be made:

1) BSWI finite element method could efficiently reduce the number of DOFs by about 3/4, but the time interval is still not good enough for an engineering problem.

2) As FFT-based BSWI FEM could improve the time interval from $0.33\mu s$ to $3.3\mu s$, while Precise integration method could improve the time interval to $50\mu s$, which has a more than 15000% improvement than traditional FEM as well as BSWI FEM and a 1500% improvement than FFT-based BSWI FEM. More importantly, the time interval is limited by the time length of load, which means the time interval for PIM could be larger if the excitation signal is longer.

3) Numerical examples have been conducted to prove that Precise integration method based BSWI FEM could precisely simulate the ultrasonic wave propagation in rod and beam structures, also crack wave and direct wave could be detected when there is crack in rod or beam.

4) Numerical example is conducted for wave propagation in rod with different diameters, which shows that the amplitude would increase when the diameter of the rod goes down. For wave propagation in beam structures, the amplitude of the signal for deflection is much larger than the amplitude of rotation DOF. Besides, the crack location could influence the amplitude of crack wave, where the amplitude of crack wave is larger when then crack is located in the middle of beam.